\documentclass[debug]{rmaa}

\usepackage{xcolor} %% texto de colores
\usepackage{soul} %% cancela texto
\usepackage{float} %%forzar posicion
\usepackage{amsmath}	% Advanced maths commands
\usepackage{wasysym}
\usepackage{natbib}
\usepackage{rotating}
\usepackage{tikz}
\usepackage{paralist}

\usepackage{psfrag,color}

\usepackage[latin1]{inputenc}

\title{H$\alpha$ kinematics of the isolated interacting galaxy pair $KPG\,486$ ($NGC\,6090$).}

\author{
  M. M. Sardaneta,\altaffilmark{1} 
  M. Rosado,\altaffilmark{2}
  and M. S\'anchez-Cruces\altaffilmark{1}}

\altaffiltext{1}{Aix-Marseille Univ., CNRS, CNES, LAM, Marseille, France.}

\altaffiltext{2}{Instituto de Astronom\'ia, UNAM, M\'exico.}

\shortauthor{Sardaneta, Rosado \& S\'anchez-Cruces}
\shorttitle{H$\alpha$ kinematics of $KPG\,486$ ($NGC\,6090$).}

\fulladdresses{
\item Margarita Rosado: Instituto de Astronom\'ia,Universidad Nacional Aut\'onoma de M\'exico (UNAM). Apdo. Postal 70-264, 04510, Ciudad de M\'exico, M\'exico (margarit$@$astro.unam.mx).
\item Minerva Mu\~noz Sardaneta and M\'onica S\'anchez-Cruces: Laboratoire d'Astrophysique de Marseille. 38 rue F. Joliot-Curie, 13388 Marseille cedex 13 France (m3sardaneta$@$gmail.com).	}

\listofauthors{M. M. Sardaneta, M. Rosado, \& M. S\'anchez-Cruces}
\indexauthor{Sardaneta, M. M.}
\indexauthor{Rosado, M.}
\indexauthor{S\'anchez-Cruces, M.}

\abstract{
In optical images, the not amply studied isolated interacting galaxy pair $KPG\,486$ ($NGC\,6090$) displays similar features to the galaxy pair The Antennae ($NGC\,4038/39$). To compare the distribution of ionized hydrogen gas, morphology and kinematic and dynamic behaviour between both galaxy pairs, we present observations in the H$\alpha$ emission line of $NGC\,6090$ acquired with the scanning Fabry-Perot  interferometer, PUMA.  For each galaxy in $NGC\,6090$ we obtained several kinematic parameters, its velocity field and its rotation curve, we also analysed some of the perturbations induced by their encounter. We verified the consistency of our results by comparing them with kinematic results from the literature. The comparison of our results on $NGC\,6090$ with those obtained in a previous similar kinematic analysis made for The Antennae highlighted great differences between these galaxy pairs.
 }

\resumen{En im\'agenes del \'optico, el par de galaxias aislado no estudiado ampliamente,  $KPG\,486$ ($NGC\,6090$), muestra caracter\'isticas similares al par de galaxias Las Antenas ($NGC\, 4038/39$). Para comparar la distribuci\'on del gas ionizado, morfolog\'ia y comportamiento cinem\'atico y din\'amico entre ambos pares de galaxias, se presentan observaciones en la l\'inea de emisi\'on  H$\alpha$ de $NGC\,6090$ adquiridas con el interfer\'ometro Fabry-Perot de barrido, PUMA. Para cada galaxia en $NGC\,6090$ se obtuvieron varios par\'ametros cinem\'aticos, su campo de velocidades y su curva de rotaci\'on, adem\'as se analizaron algunas de las perturbaciones inducidas por su encuentro. Se verific\'o la consistencia de estos resultados compar\'andolos con los de la literatura. La comparaci\'on de los resultados de $NGC\,6090$ con los de un an\'alisis cinem\'atico similar previo a Las Antenas destac\'o grandes diferencias entre estos pares de galaxias.
  }

\addkeyword{galaxies: individual ($NGC\,6090$)}
\addkeyword{galaxies: interactions}
\addkeyword{galaxies: kinematics and dynamics}
\addkeyword{methods: data analysis}
\addkeyword{techniques: interferometric}

\begin{document}
% Typeset article header
\maketitle

\section{Introduction}

%\citep{salam}

Instead of evolving in isolation, galaxies are found in clusters and groups and they can interact quite strongly with their nearby companions. These interactions can have a profound impact on the properties of galaxies, resulting in intense bursts of star formation, the onset of quasar-like activity in galactic nuclei and perhaps even the complete transformation of spiral galaxies into elliptical galaxies. Studies of galaxies in the early universe show a significant fraction of interacting and merging systems, and theories of cosmological structure formation indicate that most galaxies have had some form of strong interaction during their lifetime. Rather than being rare events, galaxy interactions may be the dominant process shaping the evolution of the galaxy population in general  \citep{cmihos}.

The evolution and interaction of galaxies is governed by gravitational effects. Morphologically, in the interacting galaxies there are large bridges and tails, stellar bars and/or increased spiral structures and, commonly, the bodies of galaxies are distorted  \citep{schweizer}. \citet{toomre}, through numerical simulations, it was  established that the gravitational interaction with another galaxy may be the source of not only spiral structures, but also of the filamentary structure, which they called `tidal tails', thus, gravity is solely responsible for these large and thin tails and several other deformations seen in several surveys \citep[e.g.][]{arp-1966, kara, vv-2001}.

Numerical simulations show that during a merger, tidal forces from a companion galaxy trigger the formation of a bar in the disc of a perturbed galaxy \citep[][]{noguchi-1987, salo-2000i, salo-2000ii, renaud-2015}. The bar acts to trigger starburst activity by rapidly funnelling large amounts of gas to the nuclear region  \citep[][]{barnes,dinshaw, Gabbasov-2014, martinc-2016}. When a bar is clearly observable in the stellar component, the gas dynamics shows very distinctive characteristics, corresponding to elliptical and non-circular orbits, the isovelocities contour curves are deformed symmetrically, then the inclination of the central isovelocity contour curve along the minor axis is one of the main features used to identify a bar \citep{bosma, combes}.

The simplest case in the process of galaxy interactions are the isolated galaxy pairs, which are systems composed of two galaxies located closely in the space that the gravitational effect of their nearest neighbours can be neglected relative to the gravitational effects exerted between each other \citep[][]{iki-2004,rosado-2011, Gabbasov-2014}. Systematic research on double galaxies provide us with important information about the conditions of formation and properties of the evolution of galaxies \citep{kara}. 

Obtaining kinematic information from  interacting galaxies is useful for understanding the effect that the interaction can have on each of the members of the pair \citep[][]{iki-2004,iki-2007,rep}. In highly disturbed velocity fields, double nuclei, double kinematic gas components and high amplitude discrepancies between both sides of the rotation curves imply strong galaxy-galaxy interactions or mergers.  On the other hand, stellar and gaseous major axes misalignments and tidal tails suggest collisions that may not always lead to merging \citep[][]{amram-2003, torresflores}.

From the observational point of view, most of the kinematic works on interacting galaxies have been carried out using long-slit spectroscopy along certain positions \citep[e.g.][]{gunthard-2016}, restricting kinematic information from only few points on the galaxy. However, for an asymmetric perturbed system, it is important to obtain the kinematic information from large portions of the disc using observational techniques such as integral field spectroscopy with a scanning Fabry-Perot interferometer (FP). In this way, the extended kinematic information can help us determine the interaction process which has been produced on each of the members of the interacting system, in addition to the fact that, sometimes the axial symmetry of each galaxy is lost during the interaction \citep[][]{iki-2004,iki-2015, benoit-2008}.%, 

\subsection*{$NGC\,6090$}

\begin{figure*}\centering
 \includegraphics[clip=true,trim=0cm 0cm 0cm 0cm,width=\textwidth,height=!]{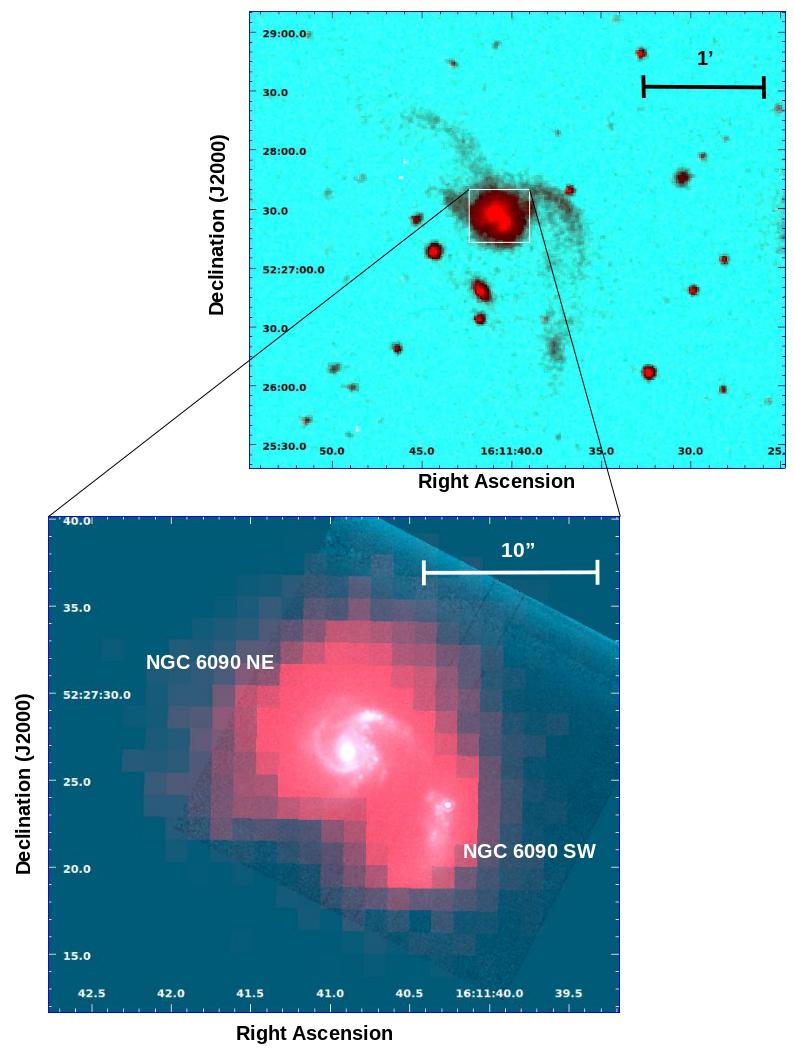}
\caption{Above: Image from the  Digitized Sky Survey (DSS) in a 4 arcmin field of view centred in the galaxy system $NGC\,6090$ over which  was  superimposed in red the H$\alpha$ direct image taken with a $6819$ \AA\@ filter with FWHM of $86$ \AA\@ obtained from the PUMA in its direct image mode at the 2.1 m telescope from the Observatorio Astron\'omico Nacional in San Pedro M\'artir (OAN-SPM), Baja California, M\'exico. Below:  2.22 $\mu$m image from the Near-Infrared Camera Multi-Object on the \textit{Hubble Space Telescope} (NICMOS/HST)  over which  was  superimposed in red the H$\alpha$ direct image from OAN-SPM in a 28 arcsec field of view centred on $NGC\,6090$.
}
\label{directa}
\end{figure*}

The isolated interacting galaxy pair $KPG\,486$ ($NGC\,6090$) has been described as a double nuclei system with an asymmetric disc and two long  tidal tails of $\sim 60$ kpc in length \citep[e.g.][]{dinshaw}.   At optical wavelengths, this galaxy system looks like the $NGC\,4038 / 39$ galaxy system (The Antennae) \citep[e.g.][]{toomre, hummel, martin, mazarella, dinshaw, bryant}.  This feature can be observed in the upper panel of Figure \ref{directa} which shows the optical image of $NGC\, 6090$ taken from the Digitized Sky Survey (DSS) in a field of view of 4 arcmin.

In radio-wavelengths $NGC\, 6090$ is described as a pair of interacting spirals separated by 0.14 arcsec with nuclei in contact and enormous curved wings \citep{martin}. Meanwhile, molecular gas appears elongated and aligned along the direction of the nuclei as a rotating disc \citep{wang-2004} or  ring \citep[][]{bryant, sugai-2000}. The position angle of that disc is $\sim 60^{\circ}$ with a major axis length of $\sim 3.4$ arcsec, the CO source appears to peak between the radio nuclei rather than on one of them \citep{bryant} and the molecular gas component does not appear to belong to any of the galaxies in the system based on the kinematics \citep{wang-2004}.

Due to the interaction evidence that $NGC\, 6090$ shows it was defined as a merger by  \citet{chisholm2015}. But also, $NGC\,6090$ has been defined as a galaxy system in an intermediate stage of merging or pre-merging too because, in addition of its two nuclei and tidal tails, the galaxy system has an identifiable bridge \citep{miralles}. In the near-infrared, the galaxies that make up $NGC\, 6090$ are seen as follows:  $NGC\, 6090\, NE$ has a distorted spiral structure and evidence of a stellar bar and $NGC\, 6090\, SW$ looks like an irregular galaxy and the position of its nucleus is still under discussion \citep[][]{dinshaw, cortijoferrero}.  In Table \ref{tab:sistema} we listed the general parameters of $NGC\,6090$ which have been reported only for a single galactic system hitherto.

To investigate the distribution of ionized hydrogen gas, morphology, and kinematic and dynamic behaviour of the galaxy system $NGC\, 6090$, we obtained a FP data cube spectrally centered on the H$\alpha$ emission line of the system. In this paper we present the analysis of the H$\alpha$ image, the velocity field and dispersion velocity map of the pair of galaxies  $NGC\, 6090$, as well as the rotation curves obtained for each member. Lastly, we comment on our results and those found in the literature and compare the kinematic characteristics of the galaxy system $NGC\, 6090$ with those of The Antennae made by \citet{amram}.

This paper is organized as follows: in \S\@ 2 there is an overview of the observational parameters and the reduction process; in \S\@ 3, we present the morphological features of $NGC \, 6090$ observed from its H$\alpha$ monochromatic and continuum maps; \S\@ 4 is devoted to the kinematic analysis made from our FP data, there are shown the derived velocity fields, the associated rotation curve of each member of the galaxy system $NGC\, 6090$ and we analyse the non-circular motions of each galaxy throughout the velocity dispersion map of the system; \S\@ 5 is dedicated to the dynamical analysis including the computation of the mass of each galaxy. A discussion is presented in \S\@ 6, and our conclusions are given in \S\@ 7.

In this paper, we considered $H_{0} = 75\,\mathrm{km\, s^{-1}\, Mpc^{-1}}$ \citep{condon} and adopted a distance of $118\, \mathrm{Mpc}$ for $NGC\, 6090$.

\newpage

\begin{table}[H]
\caption{Parameters of the galaxy system $NGC\,6090$.}
%\begin{center}
\label{tab:sistema}
 \begin{tabular}{lc}
\hline
%\multicolumn{1}{c}{Parameters} 
Parameters & $NGC\,6090$ system \\
\hline 
Coordinates (J2000) & $\alpha$ = 16$^{h}$ 11$^{m}$ 40.7$^{s}$ \\ 
 & $\delta$ = +52\arcdeg\, 27\arcmin\, 24\farcs{} $^a$\\ 
Other names & $KPG\,486$ $^{a,c}$ \\ 
& $NGC\,6090$ $^a$\\
 & UGC 10267 $^a$  \\
 & Mrk 496 $^a$ \\ 
Morphological type & G Pair$^a$ \\ 
 & Multiple galaxy$^b$  \\ 
 & Merger, double nucleus$^e$  \\ 
 & Merger$^g$\\
Mean heliocentric  & 8906$^a$ \\
radial velocity (km s$^{-1}$) & 8855$^b$ \\ 
 Distance (Mpc) & 122$^{d(*)}$  \\ 
 & 123.3$^{e(*)}$  \\ 
 & 128$^{g(**)}$  \\
 & 127.7$^{h(**)}$\\
L$_{\mathrm{IR}}$ (L$_{\odot}$) & 3$\times$10$^{11}$ $^{f}$ \\ 
m$_{b}$  & 14.36$^a$ \\ 
D$_{25 /2}$ (arcmin) & 4.36 $^b$ \\ 
Photometric nuclear separation (arcsec) &  5.4$^e$\\

 \hline
\end{tabular}
\\
\raggedright{
$^a$\citet{ned}\footnotemark[1];  $^b$HyperLeda\footnotemark[2]; \citet{leda-articulo}; $^c$\citet{kara};   $^d$\citet{condon};  $^e$\citet{bryant};  $^f$\citet{acosta-pulido};  $^g$\citet{chisholm2015};  $^h$\citet{cortijoferrero}. \\ $^{(*)}$ Distance $D= v/H_{0}$ computed using $H_{0}=75\,\mathrm{km\,s^{-1}\,Mpc^{-1}}$.\\ $^{(**)}$ Distance obtained from $z=0.02930$ with $H_{0}=70\,\mathrm{km\,s^{-1}\,Mpc^{-1}}$.
}
\end{table}

\footnotetext[1]{\textit{NASA/IPAC Extragalactic Data Base} (NED) is operated by the Jet Propulsion Laboratory, California Institute of Technology, under contract with the National Aeronautics and Space Administration. \url{https://ned.ipac.caltech.edu/}}
\footnotetext[2]{HyperLeda: \url{http://leda.univ-lyon1.fr/}}

\newpage

\section{Observations and Data Reductions}\label{obs}

Our observations to $NGC\, 6090$ were performed in July 2015 with the $f/7.5$ Cassegrain focus at the 2.1 m telescope at the Observatorio Astron\'omico Nacional in San Pedro M\'artir, Baja California, M\'exico (OAN-SPM for its acronym in Spanish) using the scanning Fabry-Perot  interferometer, PUMA \citep{rosado-1995}. PUMA is a focal reducer built at the Instituto de Astronom\'ia-UNAM used to make images and Fabry-Perot interferometry of extended emission cosmical sources. The FP used is an ET-50 (Queensgate Instruments) with a servostabilization system having a free spectral range of $19.9$ \AA\@ ($909\,\mathrm{ km\, s^{-1}}$) at H$\alpha$. The effective finesse of PUMA is $\sim 24$ which implies a sampling spectral resolution in H$\alpha$ of $0.414$ \AA\@ ($19.0 \,\mathrm{km\, s^{-1}}$) achieved by scanning the interferometer free spectral range through 48 different equally spaced channels \citep{rosado-1995}. However, due to the parallelism of the FP mirrors during the data acquisition for the observations particularly to $NGC\,6090$, the measured finesse for this data cube was $\sim 10$ which leads a sampling spectral resolution  of $0.97$ \AA\@ ($44.1\,\mathrm{km\, s^{-1}}$) at H$\alpha$.

We used a $2048\times 2048$ CCD detector with a pixel scale of $0.317\,\mathrm{arcsec}$ and we set a $4\times 4$ binning to enhance the signal, obtaining a $512\times 512$ pixel window inside of a field of view of 10 arcmin. So, we obtained a final spatial sampling per pixel of $1.27\,\mathrm{arcsec}$ which is slightly better than the average seeing of $1.67\,\mathrm{arcsec}$. To isolate the redshifted H$\alpha$  emission line, it was utilised an interference filter centred at $6819$ \AA\@ with $\mathrm{FWHM}$ of $86$ \AA\@. Thus, we obtained an H$\alpha$ data cube with a total exposure time of 72 minutes ($90\, s$ per channel).

 Then, in order to avoid phase wavelength dependence of the FP layers is necessary to calibrate the data cube in wavelength. So, we used a Ne lamp whose line at $6717$ \AA\@ is close to the redshifted nebular wavelength to obtain a calibration cube at the end of the observation. The parabolic phase map was computed from the calibration cube in order to obtain the reference wavelength for the line profile observed inside each pixel. The instrumental and observational parameters are listed in Table \ref{tabla-parametros-obs}.

\footnotetext[3]{\url{http://cesam.lam.fr/fabryperot/index/softwares} developed by J. Boulesteix.} 
\footnotetext[4]{IRAF: \textit{`Image Reduction and Analysis Facility'}\\ \url{http://iraf.noao.edu/}}

For the data reduction and analysis we used the ADHOCw\footnotemark[3] software to make the standard corrections on the cube: removal of cosmic rays and bias subtraction, subtraction of  the OH sky lines at $6828.5$ \AA,  $6842.2$ \AA\@ and $6863.9$ \AA\@ \citep[][]{ chamberlain, krassovsky} and the application of a spectral Gaussian smoothing with $\sigma=57\mathrm{\, km\,s^{-1}}$ on the data cube. We applied IRAF\footnotemark[4] tasks as  \textsc{imexamine} to determine the average seeing of our data and \textsc{mscsetwcs}  to add the World Coordinate System (WCS) to the resulting maps of $NGC\,6090$. To complete the data analysis we used our own Python scripts.

The FP scanning process allows us to obtain a flux value in arbitrary units at pixel level for each of the 48 scanning steps. The calibration in wavelength was fixed for each profile at each pixel using the calibration data cube. Thus, the intensity profile found throughout the scanning process contains information about the monochromatic emission (ionized gas emitting at H$\alpha$), the continuum emission of the object, as well as the velocity of the ionized gas. The computation of the image of the continuum was made considering the average of the 3 lowest intensities of the 48 channels of the cube  \citep[e.g.][]{vollmer-2000, iki-2004, rep, margarita-2013, nelli-2018}. For the monochromatic image, the intensity of the H$\alpha$ line was obtained by integrating the maximum value of the line profile for each pixel.

ADHOCw calculates  the radial velocity fields using the barycentre of the profile of the H$\alpha$ line for each pixel with an accuracy in central velocities of $\pm 5\,\mathrm{km\,s^{-1}}$. We masked the velocity field including only a rectangular area that contains the part of the radial velocity map corresponding to $NGC\, 6090$. Then, we superimposed the radial velocity profiles on the resulting velocity field and masked the radial velocity field based on the signal-to-noise ratio of each pixel, whose values are in the range from $24$ in the centre of the galaxy system until  $4$ in the very external pixels.

In addition to the data cube, we obtained a direct H$\alpha$ image of $NGC\, 6090$ having the FP outside the optical path and with the same filter with exposure time of 90 seconds. %The average seeing of the H$\alpha$ image is 1.67 arcsec.

\begin{table}[H]
\centering
\caption{Instrumental and observational parameters}
\label{tabla-parametros-obs}
\begin{tabular}{lc}
\hline
Parameter & Value \\ 
\hline
Telescope & 2.1 m (OAN, SPM) \\ 
Instrument & PUMA \\ 
Detector & Site3 CCD \\ 
Detector size  & $2048\times 2048$ pix\\ 
Image scale (binning $4\times 4$) & 1.27 arcsec/pix \\ 
Scanning FP interferometer & ET-50 \\ 
Finesse & $\sim 10$\\
FP interference order at H$\alpha$ & 330 \\ 
Free spectral range at H$\alpha$  & $19.9$ \AA\@ (909 $\mathrm{km\,s^{-1}}$) \\ 
Spectral sampling resolution at H$\alpha$  &  0.97 \AA\@  (44.1 $\mathrm{km\,s^{-1}}$)\\ 
Interference filter  & $6819$ \AA\@ ($\mathrm{FWHM}=86$ \AA) \\ 
Total exposure time  & 72 min \\ 
Calibration line & $6717$ \AA\@ (Ne)\\
Average seeing & 1.67 arcsec \\
\hline
\end{tabular}
\end{table}

\newpage

\section{Monochromatic and Continuum images}
\label{3.1}

On the DSS optical image of $NGC\, 6090$ (upper panel of Figure {\ref{directa}}) we superimposed the direct H$\alpha$ image obtained with PUMA. The ionized gas emitting at H$\alpha$  that we detected was found within the central area of the object, which has 20.32 arcsec per side, i.e. 16 pixels per side, where $1\,\mathrm{pix} = 1.27\,\mathrm{arcsec} = 0.726\,\mathrm{ kpc}$. The mean heliocentric radial velocity  that we determined in this work to the galactic system $NGC\,6090$ is 8885 $\mathrm{km\,s^{-1}}$. We did not detect ionized gas emitting at H$\alpha$ in the tidal tails (antennae), so we limited our kinematic analysis to the central area of $NGC\, 6090$ without taking into account the antennae. In the bottom panel of Figure \ref{directa} we show the $NICMOS/HST$ 2.22 $\mu$m image whose 22 arcsec field of view matches with our results.

\begin{sidewaysfigure}
%\begin{figure*}
\centering
 \includegraphics[clip=true,trim=0cm 0cm 0cm 0cm,width=\textwidth,height=!]{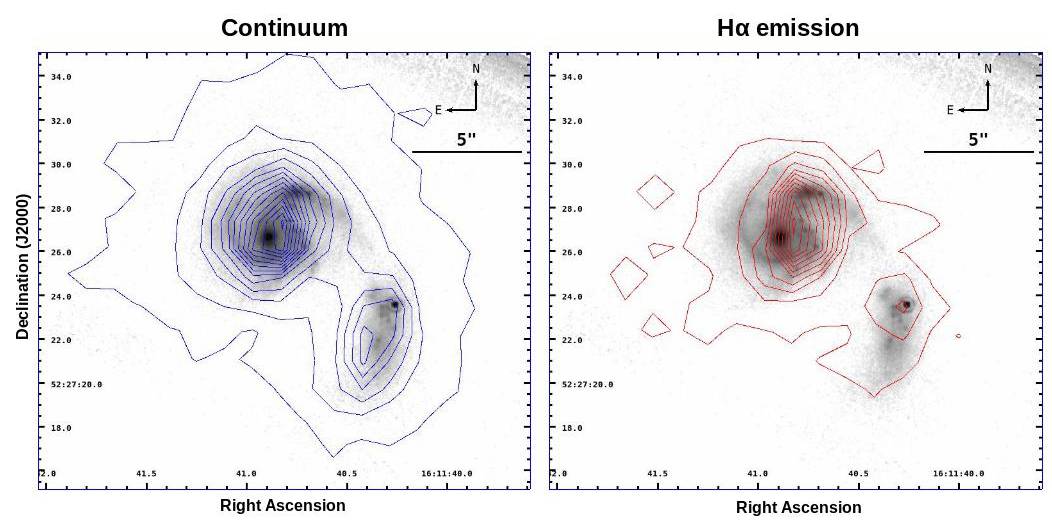}
\caption{Left: contours of the continuum linearly  spaced from 1500 to 10000 ADU obtained from the H$\alpha$ data cube from the scanning FP interferometer, PUMA overlaid on the NICMOS/HST 2.22 $\mu$m image  in a 22 arcsec field of view centred on $NGC\,6090$. Right: contours of the H$\alpha$  monochromatic image linearly spaced from 4000 to 240000 ADU obtained with PUMA  overlaid on the NICMOS/HST 2.22 $\mu$m image in a 22 arcsec field of view centred on $NGC\,6090$.
}
\label{nicmos-puma}
%\end{figure*}
\end{sidewaysfigure}

To have a sight about the morphology of $NGC\,6090$ that the PUMA data show, the contours of the H$\alpha$ continuum (Figure \ref{nicmos-puma} left panel) and monochromatic (Figure \ref{nicmos-puma} right panel) emissions  of $NGC\, 6090$ were superimposed on the $NICMOS/HST$ image.

 The coordinates of the continuum maximum of $NGC\,6090\,NE$ are $\alpha\mathrm{_{J2000}}=16^{h}11^{m}40.8^{s}$, $\delta\mathrm{_{J2000}}=+52\arcdeg\,27\arcmin\,27\farcs{}32$, and the maximum H$\alpha$ monochromatic emission is in $\alpha\mathrm{_{J2000}}=16^{h}11^{m}40.8^{s}$, $\delta\mathrm{_{J2000}}=+52\arcdeg\,27\arcmin\,27\farcs{}0$.  For $NGC\,6090\,NE$, we can observe that the ionized gas emitting at H$\alpha$ has greater presence towards the west, i.e. towards its companion galaxy, while, the isophotes of the continuum show that the stellar population has greater presence towards the east. In addition, in $NGC\,6090\,NE$ we can observe that the isophotes of the continuum are symmetrical with respect to the  photometric position angle (PA), which is the angle of the major axis of the continuum with respect to the north in counter-clockwise and whose value is $\simeq 0^{\circ}$ in this case. The pattern of the isophotes suggest the likely presence of a stellar bar. 
 
For the galaxy $NGC\,6090\,SW$, the coordinates of the photometric maximum are $\alpha\mathrm{_{J2000}}=16^{h} 11^{m} 40.4^{s}$, $\delta\mathrm{_{J2000}}=+52\arcdeg\, 27\arcmin\, 22\farcs{}21$ and the coordinates of the H$\alpha$ monochromatic maximum are $\alpha\mathrm{_{J2000}}=16^{h}11^{m}43.3^{s}$, $\delta\mathrm{_{J2000}}=+52\arcdeg\,27\arcmin\,23\farcs{}0$. For this galaxy, we can observe that the continuum has greater presence towards the south, the opposite side of its companion, and also it is accumulated towards the opposite direction of its H$\alpha$ maximum emission.  Moreover, we can note that the H$\alpha$ monochromatic maximum is nearer of the  the knot described in the literature as the maximum of the 2.22 $\mu$m image \citep[e.g.][]{dinshaw} than the photometric center of the galaxy, what has created controversy  in the literature about the position of the nuclei of $NGC\,6090\,SW$.

\newpage

\section{Kinematic results}

\subsection{Velocity fields}\label{VF}

\begin{figure*}\centering
 \includegraphics[clip=true,trim=0cm 0cm 0cm 0cm,width=1\textwidth,height=!]{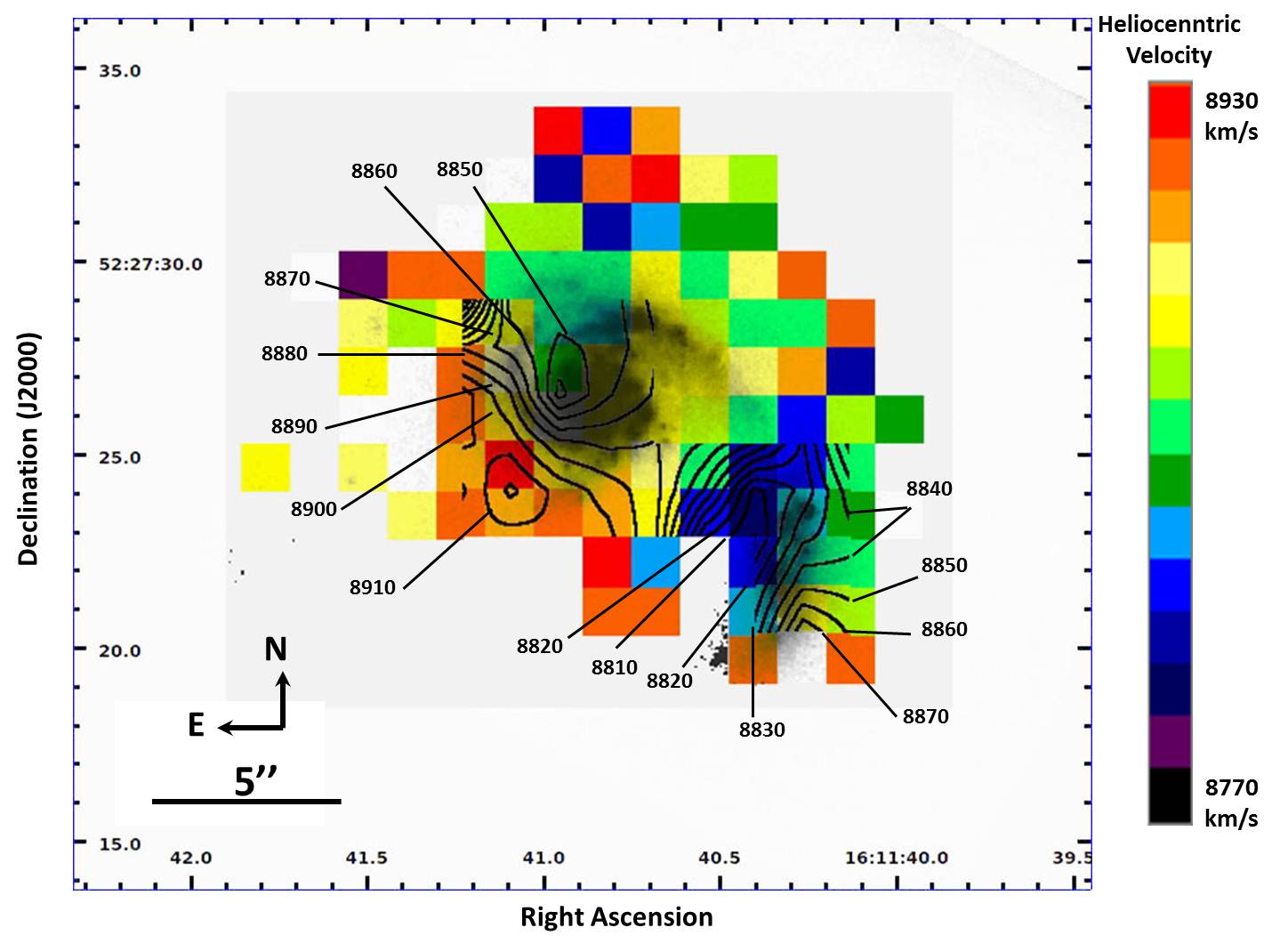}
\caption{Velocity field acquired with the H$\alpha$ data cube observation obtained with the scanning FP interferometer, PUMA, in a 26 arcsec field of view centred in $NGC\, 6090$. On the velocity field are overlaid, firstly, the NICMOS/HST 2.22 $\mu$m image and, secondly, the isovelocity contours of each galaxy which are linearly spaced by a factor of 10 starting from 8810 $\mathrm{km\, s^{-1}}$. Each isovelocity is labelled in $\mathrm{km\, s^{-1}}$ units.
}
\label{isovel}
\end{figure*}

In the previous section was shown that the ionized gas emitting at H$\alpha$  does not have the same morphology than the continuum emission, as a consequence of this phenomenon due to the interaction between the galaxies,  we do not expect that all the photometric and kinematic parameters exactly match. In Figure \ref{isovel}, which shows the velocity field of $NGC\, 6090$ where its isovelocities contour curves and the 2.22 $\mu$m image were superimposed, is possible to notice the velocity field of two independent galaxies, i.e. the velocity field of $NGC\, 6090$ presents two velocity fields ordered roughly on each one of its photometric maxima. Also, each velocity field has distortions towards the outside, mainly toward the northern side of both galaxies.

For $NGC \, 6090\, NE$, the values of the radial velocities within its ordered velocity field, which reaches a radius of  $R = 5.6\,\mathrm{ arcsec}$, are in the range from 8783 to 8919 $\mathrm{km\, s^{-1}}$. For $NGC\, 6090\, SW$, the radial velocities are in the range from 8799 to 9143 $\mathrm{km\, s^{-1}}$ within its velocity field which is ordered up to a radius of $R = 4.5\,\mathrm{arcsec}$. 
Furthermore, for $NGC \, 6090\, NE$, in Figure \ref{isovel} we observe that the isovelocities contour curves of the disc of the galaxy are not symmetric with respect to the kinematic minor axis, this feature added to the morphology signed by the continuum contours described in previous section suggests the presence of a stellar bar in agreement with the criteria proposed by \citet{bosma}.

\subsection{Rotation curves}\label{3.3}

From the morphological and kinematic features described in previous sections, we can argue that we have detected two independent galaxies inside the $NGC\,6090$ galactic system. In this kind of interacting systems is seen that the radial motions such as inflows/outflows or motions perpendicular to the galactic disc are second order effects, otherwise the discs would be destroyed. We started studying the circular motion of each galaxy in  $NGC\,6090$, which corresponds to the most basic motion that can be assumed in a disc galaxy. In this way, the rotation curve of each galaxy was obtained from the kinematic data acquired with our FP observations with the ADHOCw software, which computes the radial velocity of the galaxies in every single pixel using the barycentre of the H$\alpha$ profile observed in each pixel.

In doing so, we use the geometric and kinematic assumptions proposed by \citet{mihalas}. Supposing that each galaxy has a well-defined disc and the system rotates about an axis that is perpendicular to the galactic plane which is inclined an angle $i$ to the plane of the sky, with the polar coordinate systems ($R,\,\theta$) in the plane of the galaxy and ($\rho,\, \phi$) in the plane of the sky, locating the origin at the point where the rotation axis embeds the plane of the sky, the radial velocity ($V_{obs}$)  measured at ($\rho,\, \phi$) 
is $V_{obs}(\rho,\, \phi)= V_{sys}+V_{r}(R,\,\theta)\sin \theta\sin i  
 +V_{\Theta}(R,\,\theta)\cos \theta \sin i  + V_{z}(R,\,\theta)\cos i$, where $R^{2}=\rho^{2}(\cos ^{2} \phi+\sec ^{2}i \sin ^{2} \phi)$ and $\tan\phi =\sec i \tan \theta$, 
%%
%\begin{equation}
%\begin{split}
%V_{obs}(\rho,\, \phi)=& V_{sys}+V_{r}(R,\,\theta)\sin \theta\sin i  \\
% & +V_{\Theta}(R,\,\theta)\cos \theta \sin i \\ & + V_{z}(R,\,\theta)\cos i,
%\end{split}
%\label{mihalas1}
%\end{equation}
%
%where 
%%
%\begin{equation}
%R^{2}=\rho^{2}(\cos ^{2} \phi+\sec ^{2}i \sin ^{2} \phi)
%\label{mihalas2}
%\end{equation}
%%
%and
%%
%\begin{equation}
%\tan\phi =\sec i \tan \theta;
%\label{mihalas3}
%\end{equation}
%%
%where 
with $V_{sys}$ the mean heliocentric radial velocity of the system as a whole (that is, of its centre of mass); $V_{r}$ and $V_{\Theta}$  are the radial and tangential velocities in the plane, respectively, and $V_{z}$ is the velocity perpendicular to the plane. Given that rotation is generally the dominant form of motion in disc galaxies, we assumed $V_{r}$ and $V_{z}$ 
%in equation (\ref{mihalas1}) 
equal to zero. Moreover, under the assumption that there is axial symmetry about the centre of galaxy, $V_{\Theta}$ will depend only on $R$. Thus, the tangential velocity in the plane of the galaxy   ($V_{\Theta}$)  is obtained with the radial velocity $V_{obs_{k}}$ of the $k$th region on the galaxy in terms of observable quantities:
\begin{equation}
V_{\Theta}(R_{k})=\frac{V_{obs_{k}}-V_{sys}}{\cos\theta_{k}\sin i},
\label{mihalas4}
\end{equation}
where $R_{k}$ and $\theta_{k}$ are given in terms of the coordinates $\rho_{k}$ and $\phi_{k}$ of the $k$th galaxy region  on the sky 
which corresponds to the $k$th pixel of the velocity field that was obtained with our FP observations.

\begin{sidewaysfigure}
%\begin{figure*}
\centering 
 \includegraphics[clip=true,trim=0cm 0cm 0cm 0cm,width=\textwidth,height=!]{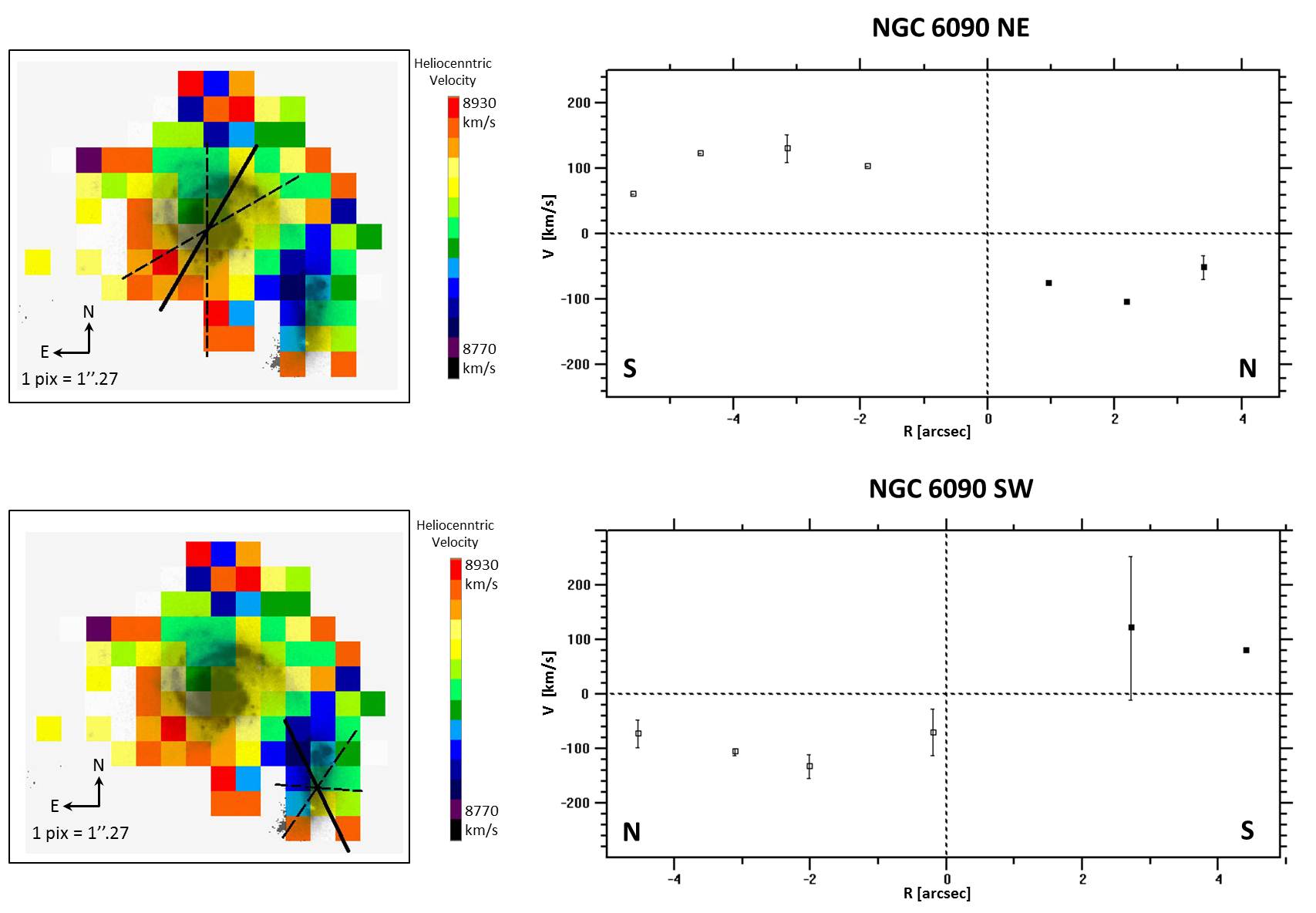}%
  \caption{
  Left panels show the velocity maps acquired with the H$\alpha$ data cube obtained with the scanning FP interferometer, PUMA, in a 22 arcsec field of view, on these maps was superimposed the NICMOS/HST 2.22 $\mu$m image as a frame of reference for the position of galaxy components in the galactic pair; then, on each map, the solid line represents the major kinematic axis which indicates the kinematic $PA$ and the dashed lines represent the sector angle that was considered for the computation of the rotation curve of each galaxy: $NGC\,6090\,NE$ above and $NGC\,6090\,SW$ below. In right panels are shown the rotation curves of each galaxy respectively obtained up to where the velocity field is ordered for each galaxy, the dispersion associated with the averaged value of pixels within the same sector is indicated with bars.
  }
  \label{crs}
%\end{figure*}
\end{sidewaysfigure}

The ADHOCw software averages the measured velocity of the pixels that are at the same distance from the kinematic center along the major axis 
indicating the dispersion associated to this average with bars on the points of the rotational observed curves. In order to avoid strong dispersions associated to the points of the rotation curves there were considered the pixels within an angular sector along the kinematic major axis  \citep[e.g.][]{iki-2004, iki-2007, rep} as is shown in left panels of Figure \ref{crs}. Then, by assuming that each galaxy has a well-defined disc,  due to its inclination it would look like an ellipse on the plane sky with apparent axis lengths $a$ and $b$, so its inclination can be computed by $i=\cos^{-1}(b/a)$. We applied the \textsc{ellipse} task of IRAF to the 2.22 $\mu$m image to estimate the inclination of the galaxies from the isophotal ellipses traced on these images. Finally, we considered that the systemic velocity of each galaxy is in agreement to the radial velocity of its own kinematic centre.

 Table \ref{tabla-CRs} contains the parameters that we used to calculate the rotation curve of each galaxy, the listed values correspond to those values with which we obtained the rotation curves from our observations, changes in these values would result in scattered points in the rotation curves instead of symmetric rotation curves.

%\newpage

\begin{table*}
%\begin{center}
\centering
\caption{Rotation curve parameters and the mass range of each galaxy in $NGC\,6090$ pair.}
\label{tabla-CRs}
\begin{tabular}{lcc}
\hline
Parameters & $NGC\,6090\,NE$ & $NGC\,6090\,SW$ \\ 
\hline
Coordinates (J2000) & $\alpha= 16^{h} 11^{m} 40.84^{s}$ & $\alpha = 16^{h} 11^{m} 40.42^{s}$ \\ 
 & $\delta$ = +52\arcdeg\,27\arcmin\, 26\farcs{}94 & $\delta$ = +52\arcdeg\, 27\arcmin\, 22\farcs{}21 \\ 
Distance (Mpc) & 118 & 118 \\ 
Systemic velocity (km s$^{-1}$) & 8880 & 8890 \\ 
Max. Rotation Velocity (km s$^{-1}$)  & 130$\pm$5 & 132$\pm$5 \\ 
Kinematic $PA$ ($^{\circ}$) & 150$\pm$3 & 26$\pm$3 \\ 
Inclination ($^{\circ}$)  & 13.9$\pm$0.5 & 64.8$\pm$0.5 \\
Radius (arcsec) &     5.6 & 4.5 \\ 
Radius (kpc) &     3.2 & 2.6 \\
Mass ($M_{\odot}$) & 0.76$\times$10$^{10}$ to 1.26$\times$10$^{10}$ & 0.63$\times$10$^{10}$ to 1.05$\times$10$^{10}$ \\ 
\hline
\end{tabular}
%\end{center}
\end{table*}

The rotation curve of $NGC\,6090\,NE$   presented in the upper right panel of Figure \ref{crs} was obtained with pixels in the velocity field within an angular sector of $30^{\circ}$ around the kinematic major axis of the galaxy as is marked in the upper left panel of Figure \ref{crs}. The physical coordinates of the kinematic centre, derived as the position around the photometric centre at which the scatter in rotation curve is minimized, are $\alpha\mathrm{_{J2000}}=16^{h} 11^{m} 40.84^{s}$, $\delta\mathrm{_{J2000}}=+52\arcdeg\, 27\arcmin\, 26\farcs{}94$. The kinematic centre used to compute the rotation curve in this galaxy matches the photometric centre within 0.4 arcsec. Using ADHOCw software, the kinematic parameters that give us the most symmetric, smooth and low scattered curve inside of the radius $R=5.6$ arcsec are  $PA = 150^{\circ}\pm 3$, $i= 13.9^{\circ}\pm 0.5$, and $V\mathrm{_{sys}}=8880\mathrm{\, km\, s^{-1}}$. We determined that the heliocentric distance to this galaxy is $118\,\mathrm{Mpc}$ through the Hubble law \citep{hubble}.  The rotation curve shows that the north of the galaxy is blue-shifted, i.e. in the north side of $NGC\,6090\,NE$ the gas approaches us, while in the south the gas moves away. Its maximum rotational velocity reaches $V_{\Theta\,max}=130\pm 5 \,\mathrm{km\,s^{-1}}$ at 3 arcsec to the south from its kinematic center.

The rotation curve of $NGC\,6090\,SW$ (bottom right panel of Figure \ref{crs}) was obtained with pixels in the velocity field within an angular sector of $60^{\circ}$ around the kinematic major axis of the galaxy. The physical coordinates of the kinematic centre are $\alpha\mathrm{_{J2000}}= 16^{h} 11^{m} 40.42^{s}$, $\delta\mathrm{_{J2000}}=+52\arcdeg\, 27\arcmin\, 22\farcs{}21$. The kinematic centre used to compute the rotation curve in this galaxy matches the photometric centre within 0.77 arcsec. The kinematic parameters that reduce significantly the asymmetry and scatter in the rotation curve inside of the radius $R=4.5$ arcsec were, in this case, $PA= 26^{\circ}\pm 3$, $i= 64.8^{\circ}\pm 0.5$, and $V\mathrm{_{sys}}=8890\mathrm{\, km\,s^{-1}}$. Just as we calculated the heliocentric distance for $ NGC \, 6090 \, NE $, we have obtained that the heliocentric distance of $ NGC\,6090\,SW $ is $ 118\,\mathrm{Mpc} $.  
The rotation curve of $NGC\,6090\, SW$ shows that the north side is blue-shifted as well as the northeastern galaxy. The maximum rotational velocity of $NGC\,6090\,SW$ is $V_{\Theta\,max} = 132\pm 5\,\mathrm{km\,s^{-1}}$ at 2 arcsec to the north from its kinematic center.

In the rotation curve of  $NGC\,6090\,SW$  there is a point on the receding side  with high dispersion associated to the average of the radial velocities  computed using only two pixels of the velocity field, the pixel on the kinematic centre considered the framework to draw the radial velocities in the rotation curve and hence,  considered the point with the lowest radial velocity value, and the pixel at the southwest of the kinematic center with the highest radial velocity value of the velocity field of all the system. Despite of that, we decided to use this point only to get a better view of the rotation curve of this galaxy toward the south.

\subsection{Velocity dispersion  field}

\begin{figure}
\begin{center}
\includegraphics[width=\columnwidth]{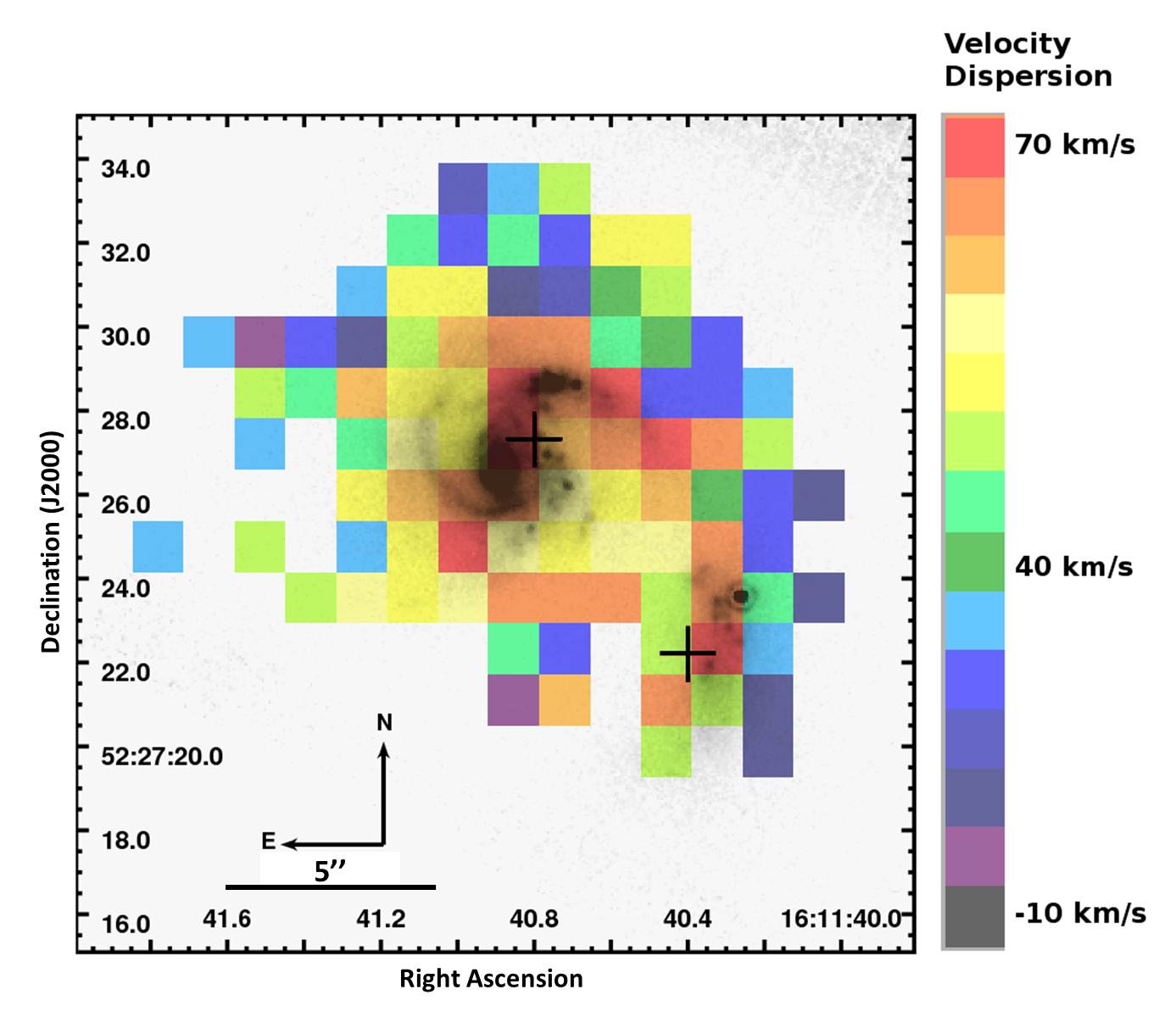}
\caption{Velocity dispersion map obtained using the PUMA in a 22 arcsec field of view centred on $NGC\,6090$ superimposed on the NICMOS/HST 2.22 $\mu$m image. Symbols `+' represent the photometric centre of each galaxy. 
}
\label{dispersion}
\end{center}
\end{figure}

Through the Full Width at Half Maximum (FWHM) of Gaussian functions fitted to the velocity profile of the H$\alpha$ line in each pixel, we obtained the velocity dispersion map of the galaxy system. Figure \ref{dispersion} displays the velocity dispersion map of $NGC\,6090$ with the NICMOS/HST image overlaid. In $NGC\,6090\,NE$ the velocity dispersion increase gradually from $\sim 45-55\,\mathrm{km\,s^{-1}}$ in the eastern side of the galaxy, to  $ 59-66\,\mathrm{km\,s^{-1}}$ in the centre, and then up to $75\,\mathrm{km\,s^{-1}}$ in its north-western spiral arm near of  what might look like a bridge between the galaxies. In $NGC\,6090\,SW$, the velocity dispersion maximum is  $85\,\mathrm{km\,s^{-1}}$ next to the position of its photometric centre, then in its surroundings the velocity dispersion range is $\sim 40 \,\mathrm{km\,s^{-1}}$ increasing to the north, where it is $\sim 69 \,\mathrm{km\,s^{-1}}$.

In the north side of galaxy system and between both galaxies, broader profiles were noticed  by double radial velocity profiles with low signal to noise ratio that we detected with PUMA mainly located in the bridge zone. Those broader profiles have been already detected by \citet{cortijoferrero} and interpreted  as a consequence of shocks in early-stage mergers. This feature of the velocity dispersion field indicates that $NGC\,6090$ has an interaction bridge between its galaxies. Those pixels with double velocity profiles are outside of the ordered velocity field of each galaxy, so that they do not contribute to any point in the computation of the rotation curves, moreover, given their low signal to noise ratio we do not consider them in this work.

%%%%
%%%

\subsection{The rotation sense of $NGC\,6090\,NE$}\label{3.4}

\begin{figure}
\begin{center}
\includegraphics[width=\columnwidth]{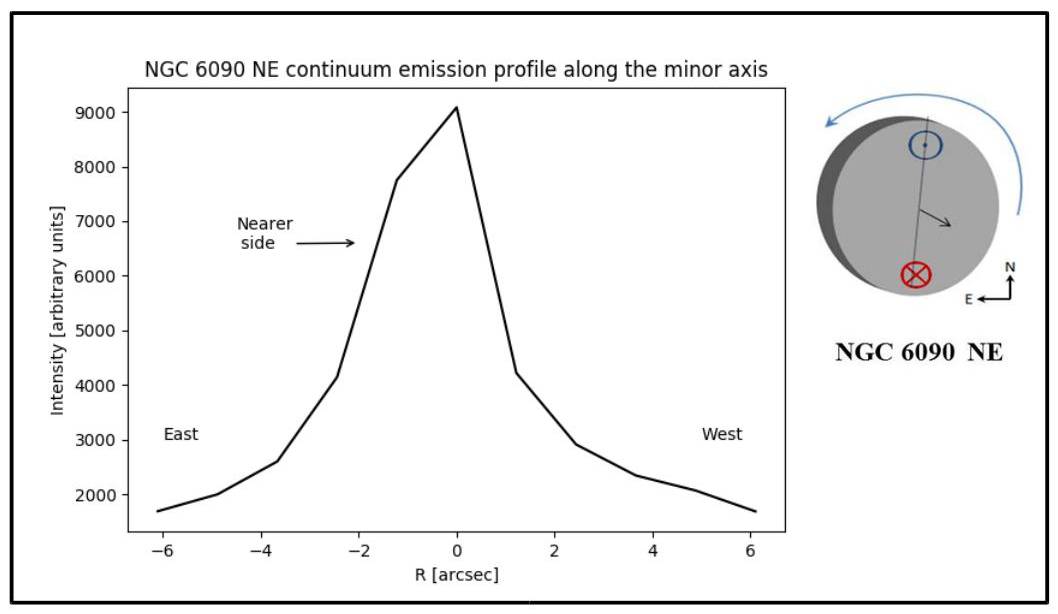}
\caption{Intensity profile of the continuum map of $NGC\,6090\,NE$ along the minor axis \textbf{centred on the galactic phometric center}. Following the gradient criterion of \citet{pasha}, the far side of the galaxy would be that one that falls more smoothly, therefore the nearest side to the observer of $NGC\,6090$ is the eastern one. Top-right panel shows a scheme of 3D orientation of $NGC\,6090\,NE$ derived from our kinematic analysis.
}
\label{brillo-NE}
\end{center}
\end{figure}
%%%

Radial-velocity measurements cannot by themselves distinguish leading and trailing spirals in thin disks. To determine whether a given galaxy leads or trails, we must determine which side of the galaxy is closer to us. Thus, to determine if the galaxies in $NGC\, 6090$ lead or trail, we followed the gradient criterion of \citet{pasha}: the apparent brightness of the nuclear region falls off unequally from the centre outward along the minor axis, the far side being the one where the surface-brightness profile declines more smoothly. It means, if the inner disc is dusty, so it absorbs a significant fraction of the starlight passing through it, then the surface brightness of the bulge at a given distance along its apparent minor axis will be lower on the near side \citep{binney}. In this way, for each galaxy in the $NGC\, 6090$ was obtained the intensity profile of the continuum emission of their minor axis in order to know the rotation sense of each one.

For $NGC\, 6090\, NE$, we extracted the intensity profile of the continuum from its minor photometric axis (see Figure \ref{nicmos-puma} in \S\@ \ref{3.1}), as it is shown in Figure \ref{brillo-NE}. This profile indicates that the nearest side to the observer is the eastern side of the galaxy. Then, considering that the rotation curve indicates that the gas on the northern side is blue-shifted and helping us with the dust lanes which trace the arms structure of this galaxy (see right panel of Figure \ref{directa}), we can infer that this galaxy has its arms rotating in a trailing direction.

On the other hand, for $NGC\, 6090\, SW$ it was not possible to obtain an accurate result through which we might determine its rotation sense due to the low number of pixels with continuum emission  along its minor axis, but in addition, the image of 2.22 $\mu$m do not mark clearly the structure of spiral arms.

%\newpage
%\clearpage

\section{Mass estimates using dynamical analysis}\label{mass}

A range of possible values for the mass of each galaxy in the $NGC\, 6090$ system was calculated using the approach described by \citet{leque}, as has been done before when studying pairs of galaxies \citep[e.g.][]{amram, iki-2007, rep}. This approach consists in calculating the mass $M (R)$ up to a certain radius where the rotation velocity $V (R)$ has been measured. The method considers two extreme cases to evaluate the mass of galaxies: the galaxy seen as a flat disc and the galaxy described as a spherical system. Thus, the mass of the spiral galaxies is in the range of $M (R) = \kappa R V^{2}(R) / G$, where $G$ is the gravitational constant and the coefficient $\kappa$ is an element of the range ($0.6,\, 1.0$), whose limiting values correspond to the galaxy being dominated by a flat disc or by a massive spherical halo, respectively.

Thus, from the $NGC\, 6090\, NE$ rotation curve, the maximum rotation velocity of this galaxy is $V_{NE} = 130\,\mathrm{km\,s^{-1}}$ which is reached at a radius of $R = 2.47\,\mathrm{arcsec}$. For this galaxy the mass range obtained is $M_{NE} = \kappa 1.26 \times 10^{10}\,M_{\odot}$. From $NGC\,6090\, SW$ rotation curve, the maximum rotation velocity is $V_{SW} = 132\, \mathrm{km\,s^{-1}}$ reached at $R = 1.97 \,\mathrm{arcsec}$. For this galaxy the mass range is $M_{SW} = \kappa 1.05 \times 10^{10} \,M_{\odot}$. Therefore, the sum of both masses in the case of both galaxies being dominated by a massive spherical halo is $M_{sph} = 2.31\times 10^{10}\,M_{\odot}$, and in the case of both galaxies being dominated by flat disc, the mass is 
$M_{flat} = 1.39\times 10^{10}\,M_{\odot}$.

When two galaxies of roughly equal mass have an encounter, their extended halos may merge to form a common halo, then the galaxies are expected to orbit in a common halo until dynamical friction and tidal interactions have removed sufficient orbital energy for the galaxies to merge \citep[e.g.][]{binney, mo}. Thus, due to the high state of interaction of the galaxies forming part of $NGC\, 6090$ it is possible infer that  they are dominated by a flat disc individually, but a further study on the dynamics of this encounter would be necessary to dismiss the probably case in which both galaxies are dominated by an spherical halo.

A second independent way of obtaining the mass value of a pair of galaxies is the method proposed by \citet{kara}, which calculates the mass of the pair from the relative orbital motion of the components of the pair. This method consists of assuming that the components of the pair move in a circular orbit with a velocity $V_{12}$ and spatial separation $r$. When $V_{12}$ and $r$ are transformed into observable quantities, they become the difference of the velocity projected in the line of view, $y = V_{12} \sin i \cos \Omega$, and the separation projected on the sky plane, $X = r(1-\sin^{2} i \sin^{2}\Omega)^{1/2}$, where $i$ is the angle between the plane of the orbit and the plane of the sky, and $\Omega$ the angle between the line of view and the line connecting the both galaxies of the pair. For a circular motion, \citet{kara} obtained the projection factor $\langle \eta\rangle = 3\pi/32$. Therefore, the total mass of the pair is:
\begin{equation}
M=\frac{32}{3\pi}\frac{X\, y^{2}}{G},
\label{kara-ec9}
\end{equation}
where $G$ is the gravitational constant. As $y$ and $X$ depend on a sinusoidal function, in order to obtain the upper limit of mass with this method and without loss of generality, for $NGC\, 6090$ we assumed that the orbital plane has an inclination of $i = 90^{\circ}$. Then, the angle between the line of sight and the line connecting the pair is $\Omega=\sin^{-1}(L/D)$, where $L = 4.1\,\mathrm{kpc}$ is the distance of the nuclei projected on the sky plane and $D = 133.1\, \mathrm{kpc}$ is the size of the line connecting the galaxies, which results in $\Omega=1.76^{\circ}$. Thus, the velocity projected on the line of view and the separation projected on the sky plane is $y\approx  V_{12}$ and $X\approx r$. Assuming a circular orbit, the total mass would be:
\begin{equation}
M_{\mathrm{orbital}}=\frac{32}{3\pi}\,\frac{r\, V\mathrm{_{12}}^{2}}{G},
\end{equation}
where $V_{12}$ is the difference between the system velocities of the galaxies, $r$ is the projected separation between the nucleus of each galaxy. For the galaxy system $NGC\, 6090$ we have, $V_{12}= 10\,\mathrm{km\,s^{-1}}$ and $r = 4.1\,\mathrm{kpc}$. Therefore, the orbital mass is $M_{orbital} = 3.23\times 10^{8}\,M_{\odot}$, which is two orders of magnitude lower than the sum of individual masses obtained with the method proposed by \citet{leque}. We think that this low value of the mass is due to the fact that the projected velocity difference $V_\mathrm{{12}}$ is quite small.

%\newpage

\section{Discussion}\label{dis}

At optical wavelengths, $NGC\, 6090$ appears as a double nuclei system with two curved antennas (tidal tails), so that this galaxy system has been described similar to $NGC\,4038 / 39$ (The Antennae)  \citep[e.g.][]{toomre, martin, mazarella, dinshaw, bryant}. With the  goal of studying kinematic similarities  between both pairs of galaxies produced by the interaction of their own galaxies, we studied the H$\alpha$ kinematics of $NGC\, 6090$ using the scanning Fabry-Perot interferometer, PUMA. The H$\alpha$ kinematics of the galaxy pair The Antennae has been studied through observations obtained with another FP by \citet{amram}. In both galaxy systems, $NGC\, 6090$ and The Antennae, no emission of ionized hydrogen was detected in their tidal tails, i.e. in both galaxy systems, ionized gas emitting at H$\alpha$  was only detected in the central zone, respectively. Therefore, \citet{amram} performed a kinematic analysis of the disc of The Antennae, as well as we have done for $NGC\, 6090$ in this paper.

\citet{amram} reported that in The Antennae, ionized gas emitting at H$\alpha$  is deficient compared with its continuum emission, but these emissions tend to grow into its farthest part to its companion. Then, the The Antennae nuclei matched unambiguously with observations at 2.2 $\mu$m. The velocity field is not ordered throughout the system, none of its component galaxies show the behavior of a rotating disc. Additionally, there were not detected pixels with double velocity profiles. Finally, using the \citet{leque} method to measure the mass of the galaxies, \citet{amram} found that the individual mass of the galaxies in The Antennae is approximately $2\times 10^{10}\,M_{\odot}$ for each one.

\begin{figure}
\begin{center}
\includegraphics[width=\columnwidth]{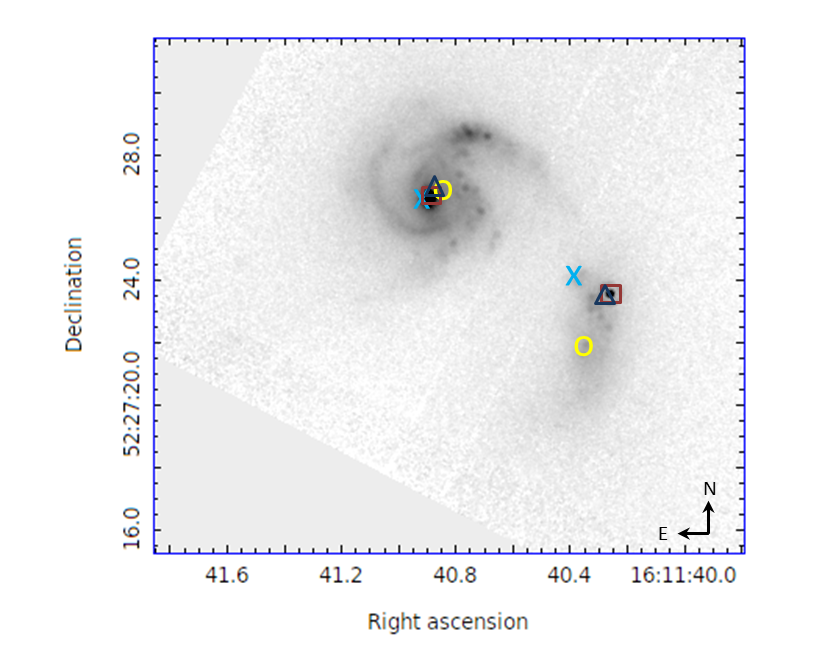}
\caption{NICMOS/HST image at 2.22 $\mu$m of a field of view of 16 arcsec centered at $NGC\,6090$. In this image we have marked the positions of the maximum emissions of the continuum (O), the H$\alpha$ monochromatic emission ($\bigtriangleup$), the radio continuum (X) and the infrared ($\Box$).
}
\label{condon}
\end{center}
\end{figure}

For $NGC\, 6090$, in \S\@ \ref{3.1},  we observed that the intensity of ionized gas emitting at H$\alpha$  is lower with respect to that of the continuum, but within each galaxy, the ionized hydrogen gas approaches to its companion, while the old stellar population is lagging in the opposite direction. The positions of the nuclei reported at 2.22 $\mu$m by \citet{dinshaw}, the HI continuum maxima reported by \citet{condon} and the monochromatic and continuum H$\alpha$ maxima that we obtained in this paper on the NICMOS/HST image, are indicated in Figure \ref{condon}.

%%%%%%
%%%

For $NGC\, 6090\, NE$, we observed that the maxima of the monochromatic and continuum of the observation made with PUMA almost completely coincide and match with the kinematic centre within 0.4 arcsec. In Figure \ref{condon} we observe that the differences between the position of the nucleus of this galaxy reported in the literature so far and our results are minimal.

However, for the galaxy $NGC\, 6090\, SW$ the maxima of the continuum and the monochromatic emissions derived from our observations of the ionized gas emitting at H$\alpha$ show that they are not almost in the same position as those in $NGC\, 6090\, NE$. This galaxy presents greater emissions of ionized hydrogen towards the north than to the south, so that the photometric maximum is located to the south while the maximum monochromatic emission is in the north. Thus, for $NGC\, 6090\, SW$, the position of the monochromatic emission maximum is 
$\alpha_{\mathrm{J2000_{H\alpha}}} = 16^{h}11^{m}40.3^{s}$, $\delta_{\mathrm{J2000_{H\alpha}}}= + 52\arcdeg\, 27\arcmin\,23\farcs{}$, and the maximum of the  continuum is  
$\alpha_{\mathrm{J2000_{cont}}} = 16^{h}11^{m}40.4^{s}$, $\delta_{\mathrm{J2000_{cont}}}= + 52\arcdeg\, 27\arcmin\,22\farcs{}$
 which match with the kinematic centre and the velocity dispersion maximum. The  maximum of the 2.22 $\mu$m emission is in   
$\alpha_{\mathrm{J2000}} = 16^{h}11^{m}40.3^{s}$, $\delta_{\mathrm{J2000}}= + 52\arcdeg\, 27\arcmin\,23\farcs{}$
 \citep{dinshaw}, and the maximum radio continuum  is in 
$\alpha_{\mathrm{J2000}} = 16^{h}11^{m}40.8^{s}$, $\delta_{\mathrm{J2000}}= + 52\arcdeg\, 27\arcmin\,27\farcs{}$
\citep[][]{hummel, condon}.

\citet{dinshaw} compared the position of the brightest  point in the 2.22 $\mu$m image  with the maximum of the radio-continuum of this galaxy, proposing that the infrared peak could be a field star. However, the emission of a field star could not be detected as the maximum of the redshifted H$\alpha$ line  of our observations of $NGC\, 6090$, therefore, it may be concluded that the  infrared maximum for the southwestern galaxy is a component itself. This possibility is supported by \citet{cortijoferrero}, who argued that the main absorption lines of their observations in the range from 3700 to 7100 \AA\@ at the knot of 2.22 $\mu$m have the same redshift of the galaxy. But also, \citet{cortijoferrero} obtained that the position of the continuum maximum is southward of the infrared peak, near from where we detected the continuum maximum, the kinematic centre and the velocity dispersion maximum of $NGC\,6090\, SW$. Thus, taking the maximum of the H$\alpha$ continuum as the nucleus of the southwest galaxy, we determined that the separation of the nuclei of the two members of $NGC\, 6090$ is $7.11\,\mathrm{arcsec}$. Assuming that the average distance of the galaxy system $NGC\, 6090$ is $118\,\mathrm{Mpc}$, the distance between its nuclei is $4.1\,\mathrm{kpc}$.

Regarding the velocity field of $NGC\, 6090$, we said in \S\@ \ref{VF} that this galaxy system has two ordered velocity fields which allowed us to compute the rotation curves of each galaxy in the $NGC\, 6090$ system. The rotation curves have higher maxima rotational velocity values  than the dispersion or residual velocity fields: the northeast galaxy has maximum rotational velocity $V_{NE} = 130 \,\mathrm{km\,s^{-1}}$ and its velocity dispersion is in the range $\sim 45 - 66\,\mathrm{km\,s^{-1}}$ inside of its disc; meanwhile $NGC\,6090\,SW$ has maximum rotational velocity $V_{SW}=132 \,\mathrm{km\,s^{-1}}$, with its velocity dispersion in the range $\sim 40 - 69\,\mathrm{km\,s^{-1}}$ inside the disc. Both galaxies reached their maximum velocity dispersion outside  the disc but even these values were still less than their maxima rotational velocities. 

Besides, with the continumm map obtained from our observations, the asymmetric isovelocities contour curves and the non-circular motions that we detected in $NGC\, 6090\, NE$,  we confirmed  the assumption about the presence of a bar structure in this galaxy made in the infrared by \citet{dinshaw} (see figure 2 of their paper) and at optical wavelengths by \citet{cortijoferrero}.

In their paper, \citet{amram} concluded that the encounter of the galaxies that make up The Antennae are in an advanced stage of interaction. In the case of $NGC\, 6090$, the features that we have discussed above about its two ordered velocity fields, continuum and monochromatic maps,  and the  behaviour of its velocity dispersion field that signals an interaction bridge between galaxies, allow us to conclude that $NGC\, 6090$ is a system in a less advanced stage of interaction than The Antennae, although both systems are in an interaction stage previous to a merger.

Regarding the mass values, Table \ref{masas} contains the values of the masses reported in the literature for the $NGC\, 6090$ system at different wavelengths compared to our results. The range of the sum of independent masses estimated in this paper with the rotation curves of each galaxy in the $NGC\, 6090$ system obtained with the method by \citet{leque} has the same order of magnitude as the masses estimated by other authors in the infrared \citep[][]{sanders, bryant, chisholm2016-1, cortijoferrero}  and radio continuum \citep{martin}, which were determined by means of the mass-to-light ratio. However, the mass of the $NGC\, 6090$ system calculated with the method proposed by \citet{kara} is two orders of magnitude smaller than those values, and it has has  an order of magnitude less as the dynamic mass estimated by \citet{bryant}.

\begin{table}
\centering
\begin{center}
\caption{Masses of the galaxy system $NGC\,6090$ found in different wavelengths in the literature compared with the ones that we found in this work.}
\begin{tabular}{lcc}
\hline
Author & Phase of the & Mass \\ 
 & IM observed & [M$_{\odot}\times$10$^{10}$] \\
\hline
\citet{sanders} & $\mathrm{H_{2}}$ & 1.4 \\ 
\citet{bryant} & $\mathrm{H_{2\,(dyn)}}$ & 0.46 \\
  & $\mathrm{H_{2\,(gas)}}$ & 2.29 \\ 
\citet{hummel} & HI & 3.9 \\ 
\citet{chisholm2016-1} & MIR & 5.02 \\
\citet{cortijoferrero} & MIR & 4.2 \\
\citet{cortijoferrero} & $3650-6950$ \AA\@ & 6.8 \\
This work & H$\alpha_{\mathrm{sum\, flat\, disc}}$ & 1.39\\
 & H$\alpha_{\mathrm{sum\, sph\, halo}}$ & 2.31\\
 & H$\alpha_{\mathrm{orbital}}$ & 0.032\\
\hline
\end{tabular}
\label{masas}
\end{center}
\end{table}

The sum of masses computed with the maximum rotation velocity coincide in order of magnitude with the masses acquired from the mass-to-light ratio at other wavelengths, even, they have the same value determined by \citet{sanders} and \citet{bryant}, either if each galaxy is dominated by a flat disc or an spherical halo, which allow us to argue that the kinematic parameters that we obtained in this work are reliable. Thus, the difference of one order of magnitude with the orbital mass can be consequence to our geometrical assumptions of the movement of $NGC\,6090\,SW$ around of its companion.

Most of the numerical simulations have taken similar initial conditions to model different features of The Antennae \citep[e.g.][]{toomre, barnes-1988, teyssier-2010, renaud-2008, renaud-2015}, either its morphology or the formation of star clusters during the merger, obtaining all the times the same feature of two long tidal tails that form the antennae. Therefore, in general, the symmetry of the tidal action allows the formation of four arms if the two companions are disc galaxies, when  the masses of the two interacting galaxies are equal or of the same order, the two internal spiral arms join up to form a bridge, the two external arms are drawn into two \textit{antennae} which remain for one or two billion years \citep{combes}. This phenomenon is happening with the galaxies of $NGC\,6090$, which have similar mass values, we note the two tidal tails and the bridge between them well defined by the nortweastern arm of $NGC\,6090\,NE$.

Finally, \citet{sugai-2004} observed young starbursts occurring in regions offset from the galactic nuclei to the northwestern spiral arm of $NGC\,6090\,NE$. They inferred that if $NGC\,6090\,NE$ had its arms trailing, the molecular gas clouds in that arm were just finishing an interaction with the central molecular gas. In \S\@ \ref{3.4} we showed that the spiral arms in the  north-eastern galaxy are trailing enforcing this theory. Another likely consequence of the trailing arms of $NGC\,6090\,NE$ could be the outflow in the northeastern side of $NGC\,6090\,SW$ detected by \citet{chisholm2016} in the UV band (see figure 1 of their paper) that coincides with a region with H$\alpha$ high velocity dispersion zone on  $NGC\,6090\, SW$, which apparently could have been formed by ram pressure during the interaction.

\section{Conclusions}

In this paper we present our observations of the ionized gas emitting at H$\alpha$ of the isolated pair of galaxies $NGC\,6090$ ($KPG\,486$) carried out with the scanning Fabry-Perot  interferometer, PUMA. This emission was detected solely in the central area of the system. Through our observations we obtained the monochromatic emission (ionized gas emitting at H$\alpha$), the continuum, the velocity field and the velocity dispersion maps. The velocity field of $NGC\, 6090$ shows two regions with ordered isovelocities contour curves from which we computed the rotation curve of each galaxy of the pair.  

Using the rotation curves, we obtained the maximum rotation velocity, the mass of each galaxy, and the mass of the galaxy system. The sum of the individual masses that we calculated is consistent with the masses found in the literature, obtained with the mass-to-light ratio for different wavelengths. In addition, we obtained the orbital mass of $NGC\,6090$, which differs in two orders of magnitude in relation to the sum of the individual masses. We conclude that this difference is due to projection effects.

In addition, enough emission was detected in the northeast galaxy to observe the specific characteristics therein. In this galaxy, the isophotes of the continuum show signs of a stellar bar, the velocity field shows no symmetry with respect to the minor axis, which in turn suggest that this galaxy does have a stellar bar. And, it was also possible to conclude that the arms of $NGC\, 6090\, NE$ galaxy rotate in a trailing direction.

We concluded that the photometric centre of the southwest galaxy is in a different location than those reported from radio and infrared wavelengths, which were inconsistent. Furthermore, the maximum of ionized gas emitting at H$\alpha$  and the maximum of the 2.22 $\mu$m coincide, confirming that this knot is part of the galaxy, and therefore it cannot be a field star. However, we were not able to study the morphological characteristics of the southwestern galaxy due to the size of the emission area that was detected.

As regards optical wavelength images of $NGC\, 6090$, the system is similar to The Antennae ($NGC\, 4038/39$), while also it has been defined as a merger. In this paper we have discussed the kinematic differences between these galaxy systems with observations made at the same wavelength with similar instruments, which allowed us to conclude that their differences are mainly due to the stage of interaction of the galaxies that make up these systems. Through the kinematic analysis of $NGC\, 6090$ we concluded that their galaxies are in a previous stage of merger than those galaxies in The Antennae.

%%%%%%%%%%%%%%%%%%%%%%%%%%%%
%%%%%%%%%%%%%%%%%%%%%%%%%%%%
%%%%%%%%%%%%%%%%%%%%%%%%%%%%

\section*{Acknowledgements}

Our research was carried out thanks to the Program  UNAM-DGAPA-PAPIIT IN109919.

The authors express their gratitude to the Mexican National Council for Science and Technology (CONACYT) for financing this paper through the project  CY-109919.

This paper is based on our observations performed with the scanning Fabry-Perot  interferometer, PUMA, at the Observatorio Astron\'omico Nacional on Sierra San Pedro M\'artir (OAN-SPM), Baja California, M\'exico.

We thank the daytime and night support staff at OAN-SPM for facilitating and helping to obtain our observations.

The Digitized Sky Surveys were produced at the Space Telescope Science Institute under U.S. Government grant NAG W-2166. The images of these surveys are based on photographic data obtained using the Oschin Schmidt Telescope on Palomar Mountain and the UK Schmidt Telescope. The plates were processed into the present compressed digital form with the permission of these institutions.

%%%%%%%%%%%%%%%%%%%% REFERENCES %%%%%%%%%%%%%%%%%%
%\bibliographystyle{aa}
%\bibliographystyle{plain}
\bibliography{kpg-references} 

%\bsp

\end{document}